\documentclass[fleqn,10pt]{wlscirep}

\usepackage{graphicx}
\usepackage{subfigure}
\usepackage{braket}

\title{Characterization of two distant double-slits by chaotic light second-order interference}

\author[1,2,@]{Milena D'Angelo}
\author[1]{Aldo Mazzilli}
\author[3,2]{Francesco V. Pepe}
\author[1,2]{Augusto Garuccio}
\author[4,5,$\dag$]{Vincenzo Tamma}
\affil[1]{Dipartimento Interateneo di Fisica, Universit\`a degli Studi di Bari, I-70126 Bari, Italy}
\affil[2]{Instituto Nazionale Di Fisica Nucleare, sezione di Bari, I-70126 Bari, Italy}
\affil[3]{Museo storico della fisica e centro studi e ricerche ``Enrico Fermi'', I-00184 Roma, Italy}
\affil[4]{Institut f\"{u}r Quantenphysik and Center for Integrated Quantum Science and Technology (IQ\textsuperscript{ST}), Universität Ulm, D-89069 Ulm, Germany}
\affil[5]{Faculty of Science, SEES, University of Portsmouth, Portsmouth PO1 3QL, UK\footnote{Present address}}
\affil[@]{milena.dangelo@uniba.it}
\affil[$\dag$]{vincenzo.tamma@port.ac.uk}

\begin{abstract}
We present the experimental characterization of two distant double-slit masks illuminated by chaotic light, in the absence of first-order imaging and interference.
The scheme exploits second-order interference of light propagating through two indistinguishable pairs of {\it disjoint} optical paths passing through the masks of interest. 
The proposed technique leads to a deeper understanding of biphoton interference and coherence, and opens the way to the development of novel schemes for retrieving information on the relative position and the spatial structure of distant objects, which is of interest in remote sensing, biomedical imaging, as well as monitoring of laser ablation, when first-order imaging and interference are not feasible.
\end{abstract}
\begin{document}

\flushbottom
\maketitle

\thispagestyle{empty}

\section*{Introduction}

In the mid fifties Hanbury-Brown and Twiss (HBT) proposed to measure the angular dimension of stars by retrieving second-order interference in the absence of first-order interference (hence, coherence) \cite{HBT1, HBT2}. The debate concerning the interpretation, and even the correctness, of HBT's predictions was quite intense due to the counterintuitive aspects related with the second-order interference arising from intensity correlation measurements \cite{brannen1956,purcell1956}. In fact, HBT intensity interferometry imposed a deep change in the concept of coherence, and triggered the development of quantum optics \cite{Glauber1963, GlauberLecture}. 

In particular, the second-order correlation measurement at the heart of HBT effect has been the working tool of all entanglement-based protocols, from Bell's inequality tests \cite{Bell} to quantum-enhanced technologies: quantum imaging and lithography \cite{imagingentangled,lithography1,lithography2,mchekhova}, information \cite{qinfo1,qinfo2,tammalaibacher2014,laibachertamma2015,tamma2014sampling}, and teleportation \cite{teleportation}. Interestingly, starting from the early 2000s, many of these effects have been replicated by exploiting the correlation of chaotic light \cite{lee2002experimental,Bennink,PhysRevA.92.043831,valencia2005two,oppel2012superresolving,PhysRevA.90.063836}. This development was enabled by the discovery that the spatio-temporal correlation exploited by HBT is not a peculiarity of the far-field of the chaotic source, but already exists in its near-field \cite{GIchaotic}. Similar to HBT, all such schemes lead to the observation of second-order interference in the absence of first-order interference. Their common element is that, given two separate detectors placed in ${\bf r}_1$ and ${\bf r}_2$, second-order interference occurs between the indistinguishable alternatives: 1) Light from point A of the source is detected in position ${\bf r}_1$, and light from point B of the source is detected in position ${\bf r}_2$; 2) light from point A of the source is detected in position ${\bf r}_2$, and light from point B of the source is detected in position ${\bf r}_1$ \cite{scully}. Most important, for second-order correlation measurement to give non-trivial results (HBT correlation peak, ghost image, ghost interference, etc.), ${\bf r}_1$ and ${\bf r}_2$ must fall {\it within} both the coherence length and the coherence area of the source. 

Recently, a novel scheme has been proposed where second-order interference is predicted to occur between light propagating through two paths that fall {\it outside} the coherence length of the source \cite{Tam-Sei}. Each interfering path is made of two disjoint, but correlated, optical paths, going from the source to a distant detector after passing though a specific arm (long or short) of an unbalanced interferometer. The unbalancing between the long and the short arm of the interferometers is such that no first-order interference exists 
at each detector, separately. However, a counterintuitive second-order interference between light propagating through the two pairs (long-long and short-short) of disjoint optical paths, 
is predicted to appear by measuring the correlation between the intensity fluctuations at the two detectors. The novelty here is that second-order interference in the absence of first-order interference is enabled by {\it a single choatic source}; in fact, in all previous schemes, second-order interference without first-order inteference relied on multiple incoherent sources \cite{mchekhova,PhysRevA.92.043831,epl68.618,pra.83.062111,prl109.233603}. 
Similar to HBT interferometry, this novel interference phenomenon
leads to a deeper understanding 
of quantum optics, and has the potential to give rise to a new research area involving both  theoretical and experimental developments. In fact, this effect,  recently described also in the spatial domain \cite{cassano}, 
enables sensing applications \cite{Tam-Sei,cassano}, 
as well as the simulation  of a C-NOT gate with a single chaotic source \cite{Tam-Sei,cassano,shih_cnot}.

In this paper, we experimentally demonstrate the spatial interference effect and the sensing technique described in Ref.\cite{cassano}  to monitor the transverse position and the spatial structure of two distant double-slit masks. The scheme is enabled by the possibility of exploiting and manipulating the relative phase characterizing the indistinguishable pairs of disjoint optical paths \cite{Tam-Sei,cassano}. Such a phase is set to zero in the case of a C-NOT gate simulation \cite{Tam-Sei,cassano,shih_cnot}, while, here, it is fully exploited for a remote sensing application \cite{cassano}. 

The sensing protocol implemented here may find applications in all those contexts where first-order imaging and interference cannot be used  for monitoring remote objects (e.g., remote sensing and biomedical imaging), as well as objects immersed in noisy environments (e.g., laser ablation).  From the fundamental point of view, the interference phenomenon at the heart of the present scheme deepens our understanding of higher-order coherence and correlation, and may lead to applications in high-precision metrology and information processing \cite{PhysRevA.90.063836,oppel2012superresolving,PhysRevA.92.043831,valencia2005two,lee2002experimental,cerf1998optical}, as well as in the development of novel optical algorithms \cite{tamma_analogue_2015-1,tamma_analogue_2015,tamma2011factoring,tamma2012prime,woelk2011factorization}.

\section*{Results}

\begin{figure}
\centering
\includegraphics[width=0.75\textwidth]{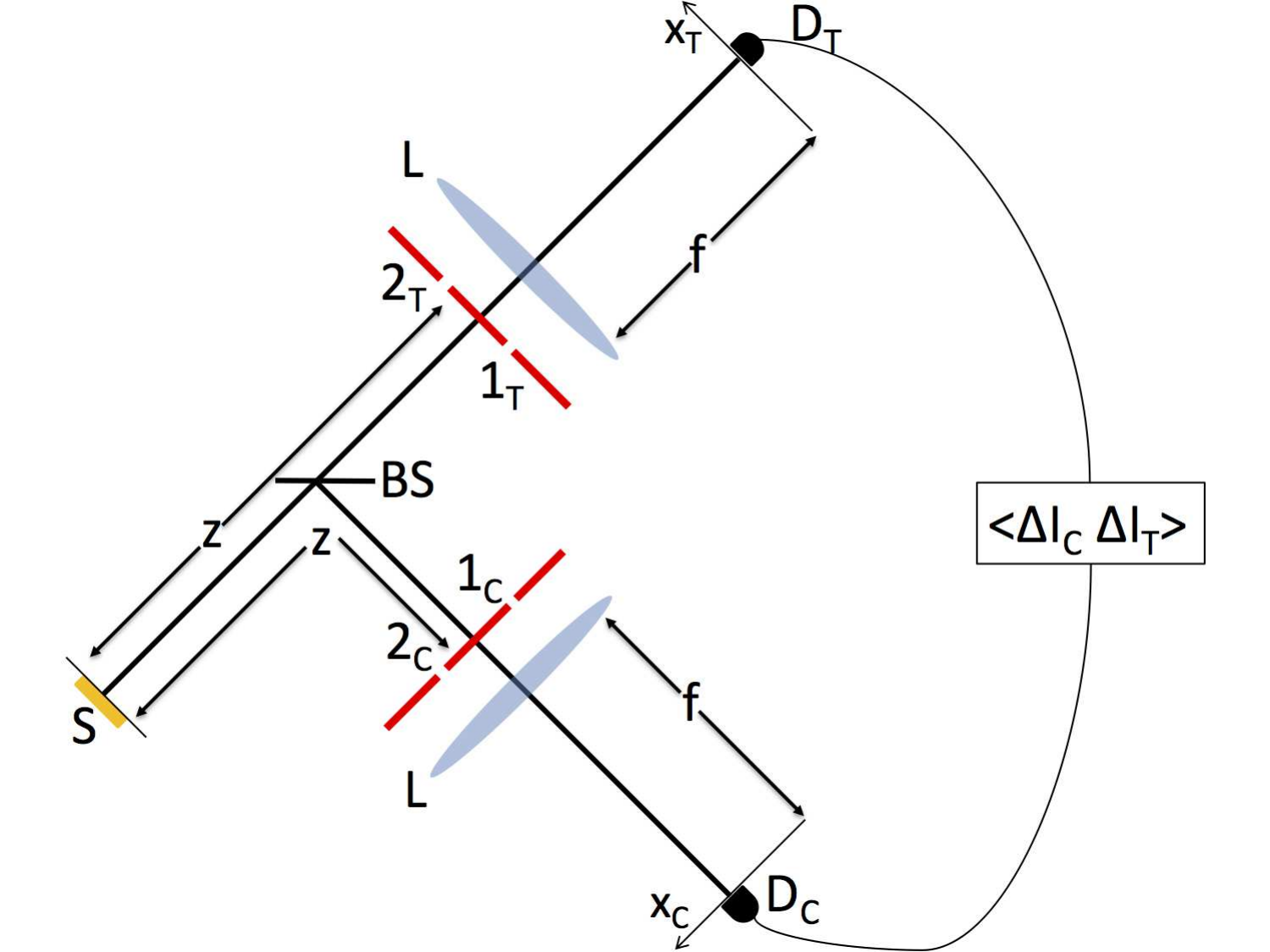}
\caption[]{Schematic representation of the experimental setup for sensing the transverse position and the spatial structure of two distant double-slit masks by second-order interference of chaotic light. See text for more details.} \label{fig:setup}
\end{figure}

As reported in Fig.~\ref{fig:setup}, we implement the spatial counterpart of the scheme proposed in Ref. \cite{Tam-Sei}, with the two Mach-Zehnder interferometers replaced by two Young (double-slit) interferometers \cite{cassano}. In particular, two double-slit masks are placed at the same distance $z$ from the source ($S$) in the transmission and reflection ports of a balanced beam splitter (BS), illuminated by chaotic light. Point-like detectors $D_C$ and $D_T$ are placed in the focal planes of the lenses ($L$) mounted behind each mask. The center-to-center distance between the slits $1_j$ and $2_j$ (with $j=C,T$) in each mask is {\it larger} than the transverse coherence length of the chaotic source; hence, no first-order interference can be observed. Let us indicate with $p^i_j$ the optical path connecting the source with the detector $j=C,T$ by going through slit $i=1,2$. We shall experimentally demonstrate that, despite paths $p^1_j$ and $p^2_j$ (with $j=C,T$) are distinguishable, second-order interference occurs between the indistinguishable pairs of {\it disjoint} optical paths  $p^{11}_{CT}=(p^1_C,p^1_T)$ and $p^{22}_{CT}=(p^2_C,p^2_T)$. In fact, interference can be retrieved at second order by measuring correlation between the intensity fluctuations detected by $D_C$ and $D_T$, namely, $\braket{\Delta I_C \Delta I_T}$, where $\Delta I_j = I_j - \braket{I_j}$, with $j=C,T$ and $\langle\dots\rangle$ denoting quantum ensemble average. Such an intriguing interference is shown experimentally to enable the sensing of both the relative transverse position and the spatial structure of the distant masks.

Let us look in more detail to the experimental setup. The chaotic light source is made of a single-mode laser diode with wavelength $\lambda=980\, \mathrm{nm}$ and power $P=300 \,\textrm{mW}$, and a rotating ground-glass disk. The distance between the source and each double-slit mask is $z=(70 \pm 5) \,\mathrm{mm}$, and the transverse coherence length of the source, on the plane of the masks, is measured to be $\ell_{coh} = (0.55 \pm 0.03) \,\mathrm{mm}$. To ensure the absence of first-order interference, we have used double-slit masks with center-to-center distances $d_j=x_{2_j}-x_{1_j}$ (with $j=C,T$) slightly larger than the transverse coherence length of the source ($\ell_{coh}$), namely $d_T= (0.57 \pm 0.03) \,\mathrm{mm}$ and $d_C=(0.69 \pm 0.03) \,\mathrm{mm}$. Still, the relative distance between the \textit{corresponding} slits of the two masks is varied within the transverse coherence length of the source; for instance, when the masks are centered with respect to each other, we have $x_{k_C}-x_{k_T} = (d_C-d_T)/2 = 60 \,\mu\mathrm{m} \simeq \ell_{coh}/10$, with $k=1,2$. The slit width is about ten times smaller than $\ell_{coh} $, the collection lenses ($L$) behind the masks have focal length $f=200\,\mathrm{mm}$, and the detectors $D_C$ and $D_T$ are amplified photodiodes with a sensitive area delimited by $50 \,\mu\mathrm{m}$-wide slits. The detectors are AC-coupled to a fast oscilloscope and connected to a computer, where a LabVIEW program performs the correlation $< \Delta I_C \Delta I_T>$ between the fluctuations of the detected intensities.

\begin{figure}
\centering
\includegraphics[width=0.6\textwidth]{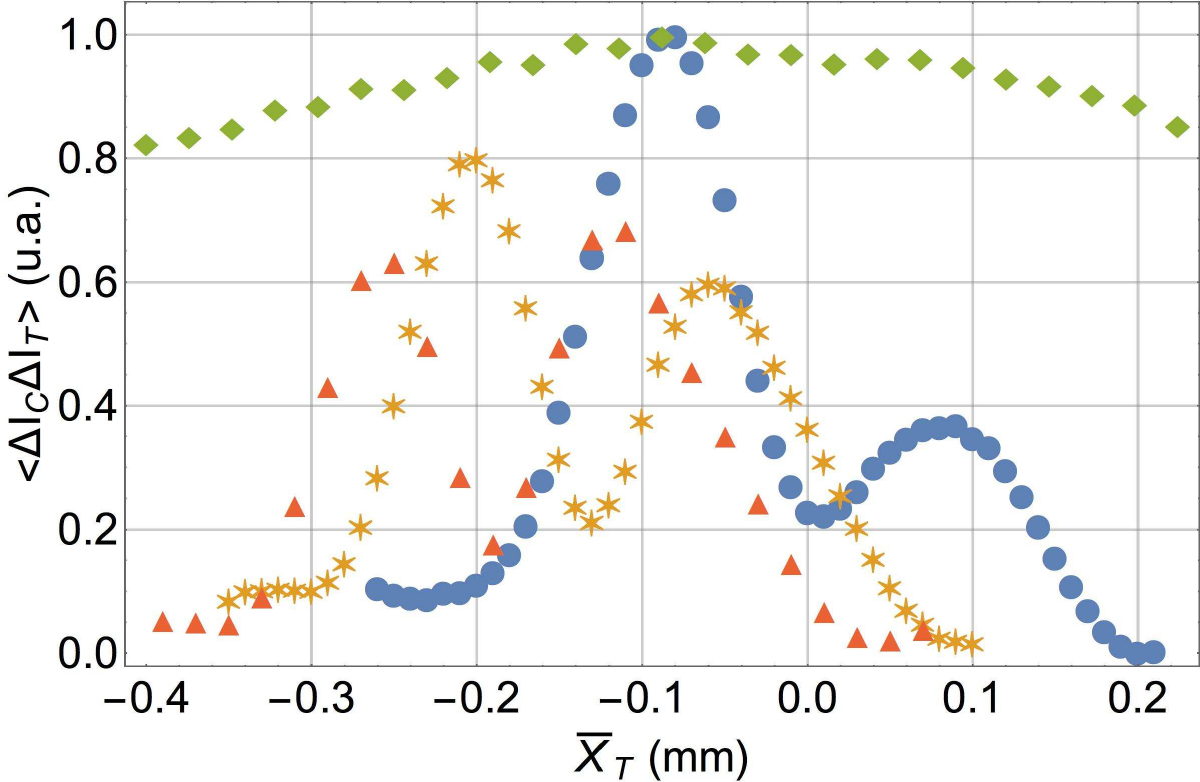}
\caption[]{Experimental demonstration of the sensitivity of second-order interference to the transverse position 
$\bar{X}_T$ of a remote double-slits ($T$) whit respect to the other ($C$). The experimental results of the correlation measurement $\braket{\Delta I_C \Delta I_T}$ are obtained by scanning the mask $T$ in the transverse plane, while keeping fixed both the mask $C$ and the two detectors $D_T$ and $D_C$. The (blue) circles, (yellow) stars and (red) triangles correspond to different positions of the fixed mask $C$. In particular, the stars and triangles are obtained after displacing mask $C$ by $\Delta \bar{X}_C=0.11\,\mathrm{mm}$ and $\Delta\bar{X}_C=0.17\,\mathrm{mm}$, respectively, with respect to its original position (circles). 
The (green) diamonds represent the normalized intensity measured by $D_T$ while scanning mask $T$ in the transverse plane; no interference effect appears at first order. Error bars are smaller than the point size for both first and second order data.
}\label{fig:mask}
\end{figure}

As reported in Fig.~\ref{fig:mask}, second-order interference between light propagating through pairs of remote slits is retrieved experimentally by moving one mask with respect to the other, while keeping both detectors $D_C$ and $D_T$ fixed. The results indicate the sensitivity of the protocol to changes in the relative transverse position of the remote masks. Interestingly, the sensitivity to the mask displacement $\bar{X}_j=(x_{1_j}+x_{2_j})/2 $ increases for masks characterized by a larger center-to-center distance $d_j$ (with $j=C,T$). 
For example, by displacing the mask $T$ with respect to the mask $C$, the expected fringe separation (see discussion below) is $\lambda z / d_T = (0.12 \pm 0.02)\,\mathrm{mm}$; this result is compatible with the measured value of $(0.15 \pm 0.02)\,\mathrm{mm}$, which has been obtained by averaging the results 
of the three sets of data reported in Fig.~\ref{fig:mask}. Notice that no first-order counterpart exists for the present results, namely, no information about the relative position of the two distant masks can be retrieved through first-order interference measurement.

Figure~\ref{fig:mask} also indicates that second-order interference is robust against misalignment of the fixed mask: Interference is not compromised when mask $C$ is displaced with respect to the optic axis. In fact, as we shall prove later, a displacement $\Delta\bar{X}_C$ of the mask $C$ is expected to shift the fringes by $\Delta\bar{X}_C d_C/d_T$; this result is experimentally confirmed by the two sets of data shown in Fig.~\ref{fig:mask}. In fact, the associated fringe displacements is measured to be $(0.14\pm 0.02) \mathrm{mm}$ for the (yellow) stars, and $(0.20\pm 0.02)\mathrm{mm}$ for the (red) triangles, both referred to the (blue) circles; this result is
 in good agreement with the theoretical prediction  
mentioned above, which gives $(0.13\pm 0.01)\mathrm{mm}$ and $(0.21\pm 0.02)\mu\mathrm{m}$, respectively. 
It is worth noticing that the envelope of the interference pattern, which is determined by the transverse coherence length of the source, shifts while moving mask $C$. In fact, we have displaced mask $C$ by a distance $\Delta\bar{X}_C$ that is not negligible with respect to $\ell_{coh} $; the envelope thus moves by $\Delta\bar{X}_C$. The modified shape of the interference fringes in the three data sets is due to the correction factor $d_C/d_T$ characterizing the displacement of the interference pattern with respect to the displacement of the envelope alone. 

\begin{figure}[!t] 
\centering
\includegraphics[width=0.9\textwidth]{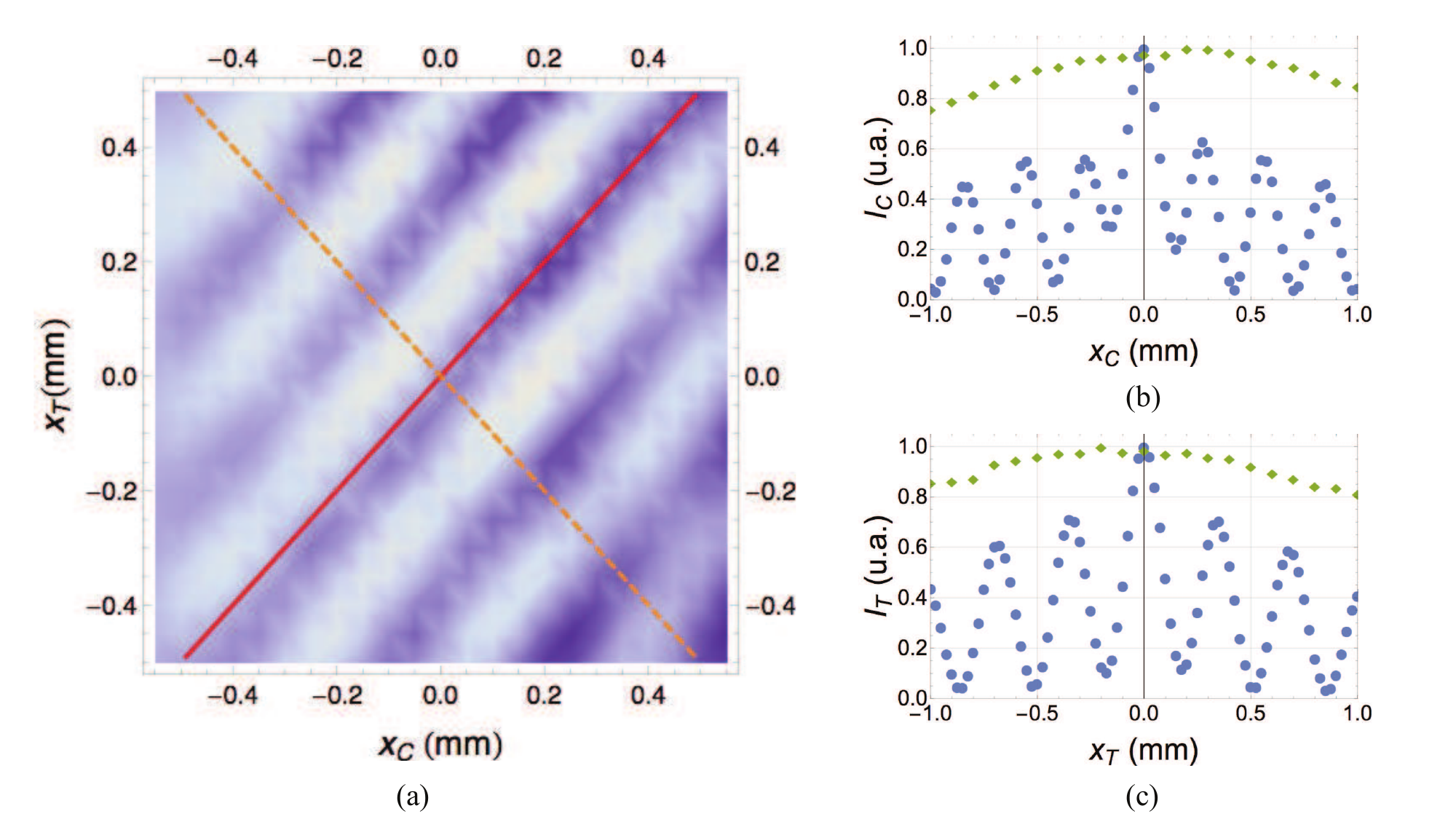}
\caption[]{Second-order interference measurement of the center-to-center separations $d_C$ and $d_T$ of the two remote double-slits masks $C$ and $T$. a) Experimental results of the normalized correlation measurement $\braket{\Delta I_C \Delta I_T}$ obtained by simultaneously scanning the transverse positions $x_C$ and $x_T$ of both detectors $D_C$ and $D_T$, while keeping both masks fixed. From the fringe periodicity along each axis $x_j$ ($j=C,T$) it is possible to retrieve the value of the center to center distance $d_j$ characterizing the corresponding mask $j$. Moreover, the periodicity of the diagonal (solid red line) and anti-diagonal (dashed orange line) interference patterns provides the values of the difference $d_C-d_T$ and the sum $d_C+d_T$ of the center-to-center distances, respectively. The tilt of the equal-phase lines in the $(x_C,x_T)$ plane is determined by the ratio $d_C/d_T$ of the center-to-center distances characterizing the two masks [see Eqs.~(\ref{eq:Delta12_phi})-(\ref{phi})]. b) Normalized intensity measured by detector $D_C$ while scanning the transverse plane $x_C$, when either laser light  (blue circles) or chaotic light (green diamonds) illuminates the mask. c) Normalized intensity measured by detector $D_T$ while scanning the transverse plane $x_T$, when either laser light  (blue circles) or chaotic light (green diamonds) illuminates the mask. Notice the absence of chaotic-light interference at first-order in both cases b) and c). Error bars are smaller than the point size for both sets of data.} \label{fig:xCxT}
\end{figure}

In Fig.~\ref{fig:xCxT}a), we experimentally demonstrate the existence of a spatial beating effect associated with the simultaneous displacement of both detectors $D_C$ and $D_T$. The fringes along the diagonal ($x_T=x_C$) and the anti-diagonal ($x_T=-x_C$) of the bidimensional plot are the typical interference fringes of a double-slit having center-to-center distance $d_C-d_T$ and  $d_C+d_T$, respectively. The correlation measurement is thus sensitive to the characteristic dimensions of both masks, even in the absence of first-order interference.  
To emphasize the pure second-order nature of the phenomenon, in Fig.s~\ref{fig:xCxT} b)-c) we show the lack of first-order interference at both detectors $D_C$ and $D_T$.  In particular, we compare the interference fringes obtained at first-order when the rotating ground-glass disk is removed and the masks are illuminated by pure laser light (blue circles), with the pattern obtained by illuminating the masks with chaotic light (orange diamonds). For chaotic light, first-order interference is washed out, as expected. However, the absence of first-order interference does not affect second-order interference, as demonstrated by Fig.~\ref{fig:xCxT} a), where second-order interference fringes at both detectors $D_C$ and $D_T$ are clearly visible along the axis $x_C$ and $x_T$, respectively. Their expected periodicity is determined by the center-to-center distance $d_j$ of the double slits (with $j=C,T$), as $\Delta x=\lambda f / d_j$, which gives $(0.34 \pm 0.02)\,\mathrm{mm}$ for mask $T$ and $(0.28 \pm 0.01)\,\mathrm{mm}$ for mask $C$. The measured periodicities of $(0.34 \pm 0.02)\mathrm{mm}$ and $(0.29 \pm 0.02)\mathrm{mm}$ are thus in excellent agreement with the theoretical predictions. 

The error on the correlation measurements has been evaluated as\cite{BoydSNR} $\sigma_{corr} /\langle\Delta I_c \Delta I_T\rangle =  (2 N \pi \ell_{coh}^2/A_{mask} )^{-1/2} \sim 10^{-3}$, where $N=10^4$ is the number of measurements, and $A_{mask}=0.46\, \mathrm{mm}^2$ is the average mask transmission area. The error on the intensity measurement is also very small: $\sigma_I /\langle I\rangle = N^{-1/2} \sim 10^{-2}$.

The presented measurement is useful for monitoring changes in both the spatial structure and the relative position of a remote mask with respect to a reference local mask; such changes may be caused by temperature variations, as well as deformations, wearing down and displacements due to interaction with the environment.

\section*{Discussion}

The above experimental results can be quantitatively understood by considering the second-order correlation function of a quasi-monochromatic chaotic light source \cite{D'AnShi}
\begin{eqnarray}\label{G2_chaot}
G^{(2)}(x_j, x_k) 
= & G^{(1)}(x_j) G^{(1)}(x_k)+|G^{(1)}(x_j, x_k)|^2,
\end{eqnarray}
where $G^{(1)}(x_j)=\braket{I_j(x_j)}$ is the average intensity at the transverse position $x_j$, and $|G^{(1)}(x_j, x_k)|^2=\braket{\Delta I_j(x_j) \Delta I_k(x_k)}$ is the correlation between the fluctuations of the intensities detected at the remote transverse positions $x_j$ and $x_k$, separately. The AC-coupling of signals from the detectors cancels the trivial contribution in Eq.~(\ref{G2_chaot}), coming from the DC-components of the detected signals, thus leaving only the interesting part of the second-order correlation $|G^{(1)}(x_j,x_k)|^2$ \cite{Scarcelli}. This is how the constant background typical of chaotic light has been removed from our experimental measurement.

Based on the position-position correlation at the heart of ghost imaging with chaotic light \cite{Scarcelli, ValScaD'AnShi, gatti2, ferri}, the fluctuation correlation measurement between the intensities detected by $D_C$ and $D_T$ is expected to yield a nonvanishing contribution for all four possible pairs of paths $p^{\alpha\beta}_{CT}=(p^{\alpha}_C,p^{\beta}_T)$, provided the relative transverse distance between \textit{each pair of remote slits} $\alpha,\beta=1,2$ is smaller than the transverse coherence length of the source, namely, $|x_{\alpha_C}-x_{\beta_T}|\lesssim \ell_{coh}$. We shall now consider the more interesting case, studied in Ref.\cite{cassano} and here implemented experimentally, in which no first-order interference exists behind each mask, which occurs whenever the distance between the two slits of a given mask is larger than the transverse coherence length of the source:
\begin{eqnarray}\label{x12_large}
|x_{1_j} - x_{2_j}| \gtrsim  \ell_{coh},
\end{eqnarray}
with $j=C,T$. We still keep the hypothesis that the corresponding slits at the two remote masks are within the transverse coherence length, that is
\begin{eqnarray}\label{x12_small}
|x_{\alpha_j} - x_{\alpha_k}| \lesssim  \ell_{coh}, 
\end{eqnarray}
with $\alpha=1,2$ and $j,k=C,T$ (with $j \neq k$). In this scenario, only the two pairs of remote paths $p^{11}_{CT}$ and $p^{22}_{CT}$ are expected to contribute to the correlation measurement, while no contribution is expected from the two pairs of paths $p^{\alpha\beta}_{CT}$ with $\alpha\neq\beta$. An interesting question is whether these contributions add coherently or incoherently, hence, whether or not they can give rise to second-order interference.

One may attempt to answer this question based on ghost imaging, where a bucket detector is placed behind the object, and a point-like detector scans the {\it imaging plane}, whose distance from the source is equal to the object-to-source distance: The scanning detector retrieves the image of the object-mask through correlation measurements \cite{Scarcelli, ValScaD'AnShi}. In ghost imaging, a double-slit is resolved only if its center-to-center distance is larger than the transverse coherence length of the source, on the object plane \cite{GI_resolution}; this is exactly the case we are considering. Ghost images are well known to be incoherent images, given by the incoherent sum of the contributions coming from each slit, separately \cite{ValScaD'AnShi}. Now, in the setup of Fig.~\ref{fig:setup}, we have made two important changes with respect to ghost imaging: 1) A mask has been placed in the ghost imaging plane of the other mask; 2) both detectors are point-like and have been moved in the far-field of the two masks. One may expect that, similar to ghost imaging, the contributions associated with the two paths $p^{11}_{CT}$ and $p^{22}_{CT}$ would add incoherently. In fact, this would effectively be the case if we had left, behind either one of the object-masks, a bucket detector. However, our fluctuation correlation measurement is performed between two \textit{point-like} detectors $D_C$ and $D_T$. As we shall show, this scheme gives rise to a coherent superposition of the two indistinguishable alternatives $p^{11}_{CT}$ and $p^{22}_{CT}$, thus leading to the observed second-order interference effect \cite{footnote1}.
This interference effect can be formally demonstrated by evaluating the spatial correlation between the intensity fluctuations $\Delta I_C(x_C)$ and $\Delta I_T(x_T)$ measured at the same time $t=t_C=t_T$, by detectors $D_C$ and $D_T$, namely
\begin{equation} \label{Delta} 
\braket{\Delta I_C(x_C) \Delta I_T(x_T)} \propto \left| \braket{\hat{E}_C^{(-)}(x_C) \hat{E}_T^{(+)}( x_T)} \right|^2
\end{equation}
where 
$\hat{E}_j(x_j)$, with $j=C,T$, is the electric field operator (in the scalar approximation) in the transverse position $x_i$, with $(\pm)$ denoting positive- and negative-frequency parts.
In the paraxial approximation, we obtain \cite{cassano}
\begin{equation} \label{Delta12_complete} 
\braket{\Delta I_C(x_C) \Delta I_T(x_T)} \propto \big{|}G^{(1)}_{1_C,1_T}(x_C, x_T) + G^{(1)}_{1_C,2_T}(x_C, x_T)+G^{(1)}_{2_C,1_T}(x_C, x_T) + G^{(1)}_{2_C,2_T}(x_C, x_T)\big{|}^2,
\end{equation}
where $G^{(1)}_{\alpha_C,\beta_T}$ indicates the contribution to the correlation measurement coming from the optical path $p^{\alpha\beta}_{CT}$, linking the remote slits $\alpha_C = 1_C, 2_C$ and $\beta_T = 1_T, 1_T$ by passing through the source. For simplicity, we shall neglect the slit width.
In this hypothesis, the first-order correlation function associated with the generic {\it disjoint} optical path $p^{\alpha\beta}_{CT}$ is given by \cite{cassano}:
\begin{eqnarray} \label{G1_ij} 
G^{(1)}_{\alpha_C,\beta_T} (x_C, x_T) = \exp\left[ \mathrm{i}(\varphi_{\beta_T} - \varphi_{\alpha_C})\right] \mathcal{S}\!\left(\frac{x_{\beta_T}-x_{\alpha_C}}{\lambda z}\right),
\end{eqnarray}
where $\mathcal{S}$ is the Fourier transform of the source intensity profile, and
\begin{eqnarray}\label{B}
\varphi_j = \frac{2 \pi }{\lambda}\left( \frac{x_j^2}{2z} - \frac{x_d x_j}{f}  \right),
\end{eqnarray}   
with $j = \alpha_C , \beta_T$. The dependence of the result of Eq. (\ref{G1_ij}) from the Fourier transform of the light source profile, evaluated in the object-mask plane, explicitly indicates the important role played by the conditions in Eqs.~(\ref{x12_large})-(\ref{x12_small}): They imply $G^{(1)}_{1_C,2_T}$ and $G^{(1)}_{2_C,1_T}$ to vanish, within a good degree of approximation, thus reducing Eq.\ (\ref{Delta12_complete}) to \cite{cassano}: 
\begin{equation} \label{Delta12} 
\braket{\Delta I_C(x_C) \Delta I_T(x_T)} \propto \big{|}G^{(1)}_{1_C,1_T}(x_C, x_T) + G^{(1)}_{2_C,2_T}(x_C, x_T)\big{|}^2,
\end{equation}
In this condition, the intensity fluctuation correlation measurement [Eq.~(\ref{Delta12})]  enables retrieving the \textit{second-order interference} between the first-order correlation functions $G^{(1)}_{1_C,1_T}$ and $G^{(1)}_{2_C,2_T}$ associated with the two disjoint optical paths $p^{11}_{CT}$ and $p^{22}_{CT}$, respectively. Such second-order interference between paths $p^{11}_{CT}$ and $p^{22}_{CT}$ is quite counterintuitive, considering the absence of coherence between the composing paths $p^{1}_{C}$ and $p^{2}_{C}$ (or $p^{1}_{T}$ and $p^{2}_{T}$). Interference between $p^{11}_{CT}$ and $p^{22}_{CT}$ stems from their indistinguishability, which is preserved even if the double-slit interference pattern produced by each mask, separately, is hindered by the conditions $d_j=|x_{2_j}-x_{1_j}|\gtrsim \ell_{coh}$ for $j=C,T$.

Plugging in the results of Eqs.~(\ref{G1_ij})-(\ref{B}), we shall rewrite Eq.~(\ref{Delta12}) in a more explicit form\cite{cassano}:
\begin{eqnarray}\label{eq:Delta12_phi}
\braket{\Delta I_C(x_C) \Delta I_T(x_T)}  \propto \left| \mathcal{S}\!\left(\frac{x_{1_C}-x_{1_T}}{\lambda z}\right) + \mathcal{S}\!\left(\frac{x_{2_C}-x_{2_T}}{\lambda z}\right) e^{  \imath \phi(\bar{X}_C, d_C, \bar{X}_T, d_T, x_C, x_T)}\right|^2, 
\end{eqnarray}
with
\begin{eqnarray}\label{phi} 
\phi(\bar{X}_C, d_C, \bar{X}_T, d_T, x_C, x_T) &=& \frac{2 \pi}{\lambda z}  (\bar{X}_T d_T - \bar{X}_C d_C)  - \frac{2 \pi}{\lambda f}  (x_T d_T - x_C d_C),
\end{eqnarray}
where $d_j=x_{2_j}- x_{1_j}$ is the center-to-center separation between the slits in each mask, and $\bar{X}_j=(x_{1_j}+ x_{2_j})/2$ is the center of each mask. The phase of Eq.~(\ref{phi}) is at the core of our remote sensing experiment.

In fact, the experimental results reported in Fig.~\ref{fig:mask} are related with the first term of the phase defined in Eq.~(\ref{phi}), and correspond to the situation in which both the mask $C$ and the detectors are kept fixed, while the mask $T$ is moved in the transverse plane. The results of Eq.~(\ref{eq:Delta12_phi}) and ~(\ref{phi}) thus formally demonstrate the observed sensitivity of the intensity fluctuation correlation measurement to the relative positions of the double-slit masks \cite{cassano}.

The results shown in Fig.~\ref{fig:xCxT}a) are instead related with the last term of the phase given in Eq.~(\ref{phi}). On one hand, based on Eq.~(\ref{eq:Delta12_phi}) and ~(\ref{phi}), by scanning $x_C=\pm x_T$ one can foresee the observed spatial beating effect, namely, fringes with a periodicity determined by the combination of the center-to-center distances characterizing the remote masks ($d_T \mp d_C$) \cite{cassano}. On the other hand, Eqs.~(\ref{eq:Delta12_phi})-(\ref{phi}) indicate the possibility of retrieving, at second-order, the standard Young-type interference pattern associated with each double-slit mask; the pattern can be obtained by scanning only one detector along the transverse direction $x_j$ (with $j=C,T$), and is centered in $x_{i}=x_{j} d_{j}/d_{i}$, with $i,j=C,T$ and $i \neq j$. This formally demonstrates the sensitivity of the intensity fluctuation correlation measurement to the spatial structure of both double-slit masks. Interestingly, the scheme is feasible at arbitrary distances between the masks, since it only requires the chaotic source to be placed at the same optical distance from each mask.

In summary, the sensing technique implemented here enables retrieving information about the spatial structure and position of distant masks despite the absence of first order coherence. When first order interference exists, the information about the two masks may be encoded in the second order interference resulting from all four pairs of paths\cite{footnote2};
 however, decoding the relevant information from the measured correlation is not trivial in such situations and will be studied elsewhere.

\section*{Conclusion}
In conclusion, we have experimentally demonstrated the possibility of exploiting an intriguing second-order interference effect of chaotic light for monitoring the transverse position and the spatial structure of two distant double-slit masks. In fact, we have shown that the experimental results are due to the interference between two effective optical paths $p^{\alpha\alpha}_{CT}$, with $\alpha=1,2$, each one made of the disjoint, but correlated, paths $p^{\alpha}_C$ and $p^{\alpha}_T$ associated with the remote slits $\alpha_C$ and $\alpha_T$, respectively (see Fig.~\ref{fig:setup}). Interestingly, such interference occurs even if the two slits $1_i$ and $2_i$ of both masks $i=C,T$ are outside the coherence area of the source, so that no first-order interference exists. Interference is recovered at second-order because the planar distance between the remote slits $\alpha_C$ and $\alpha_T$ is smaller than the transverse coherence length of the source.

The generalization of our results to more general objects may lead to the development of novel methods for retrieving information about the position and the spatial structure of two distant objects \cite{cassano}. Our results are thus of potential interest for applications in imaging, sensing and metrology, also in the presence of noise. In addition, the extension of this technique to correlation measurements of order larger than two \cite{Tam-Sei} might be used both to spatially resolve a larger number of remote objects and to improve the imaging precision \cite{oppel2012superresolving,PhysRevA.92.043831,classen2016superresolving}. Future research will also aim at replacing the chaotic illumination with both entangled light sources (e.g., spontaneous parametric down-conversion), which would enable sub-shot noise sensing \cite{sub_shot}, and atomic systems for fundamental tests with quantum matter \cite{interferometryBEC}.

\section*{Acknowledgements}

V.T. acknowledges Michele Cassano for useful discussions. This work has been supported by MIUR project P.O.N.~RICERCA E COMPETITIVITA' 2007-2013 - Avviso n.~713/Ric.~del 29/10/2010, Titolo II - ``Sviluppo/Potenziamento di DAT e di LPP'' (project n.~PON02-00576-3333585). M.D., A.G. and F.V.P. acknowledge support by Istituto Nazionale di Fisica Nucleare (INFN) through the project QUANTUM. V.T. acknowledges the support, during the first part of this project, of the German Space Agency DLR with funds provided by the Federal Ministry of Economics and Technology (BMWi) under grant no. DLR 50 WM 1556.

\section*{Author contributions statement}

M.D., A.G., and V.T. conceived the original idea, M.D. designed the experimental setup, M.D. and A.M. conducted the experiment, A.M. and F.P. analysed the results. M.D. wrote the paper. All authors contributed to the final version of the manuscript.

\section*{Additional information}

The authors declare no competing financial interests.


\begin{thebibliography}{10}
\bibitem{footnote1}
{Incidentally, it is known that the standard (incoherent) ghost image becomes coherent as soon as one replaces the bucket detector with a point-like detector placed in the far field of the mask \cite{coherentGI,pleno1}. The use of a bucket detector washes out the coherence, and the related interference, by averaging all interference patterns at different detection positions.}
\bibitem{footnote2}
{When first-order interference has $100\%$ visibility, the second-order correlation function in Eq. (\ref{Delta12_complete}) reduces to the mere product of the interference patterns measured at the two detectors, separately: $G^{(1)} _{1_d,2_d} \propto 1 + \cos (\varphi_{2_d} - \varphi_{1_d})$, with $d = C, T$; hence, in this case, the correlation measurement does not add any information with respect to the one already retrievable at first order.}
\end{thebibliography}
\end{document}